\documentclass[preprint,11pt]{elsarticle}
\usepackage{amsmath,color}
\usepackage{amssymb}
\usepackage{theorem}
\usepackage{amsfonts}

\usepackage{graphicx}
\usepackage{epsfig}
\begin{document}
\begin{frontmatter}
\title{ Using a priori knowledge to construct  copulas 
   }
 \author{Dominique Drouet Mari, Val\'erie Monbet}
 \address{ Universit\'e Europ\'enne de Bretagne, Lab-STICC (UMR CNRS 3192),\\
  Universit\'e de Bretagne-Sud, \\
  Centre Yves
Coppens, Campus de Tohannic, \\
F-56017, Vannes, France
}

\begin{abstract}
 Our purpose is to model the dependence between two random variables,  taking into  account  a priori knowledge on these variables. For example, in many
applications (oceanography, finance...), there exists an order relation between the two variables; when one takes high values, the other cannot take
low values, but the contrary is possible. The dependence for the high values of the two variables  is, therefore,  not symmetric. 
   However a minimal dependence also exists: low values
of one variable are associated with low values of the other variable. The dependence can also be extreme for the maxima or the minima of the two variables.

In this paper, we construct step by step asymmetric copulas with  asymptotic minimal dependence, and with or without asymptotic maximal dependence, using mixture variables to get at first  asymmetric dependence and then minimal dependence.  We fit these models to a real dataset  of sea states and compare them using Likelihood Ratio Tests when they are nested, and BIC- criterion (Bayesian Information criterion)  otherwise. 

 \end{abstract}

\end{frontmatter}

\section{Motivation }
Since the nineteen sixties and the pionnering works of Gumbel \cite{Gumbel}, Plackett \cite{Plackett}, Mardia \cite{Mardia}, the construction of bivariate distributions  with fixed margins (i.e. the construction of copulas) has interested many researchers. 

~

\noindent
Various  procedures to contruct copulas have been proposed. A fruitful method is to construct dependence by  mixing with respect to a third random variable, called a frailty variable (see Clayton \cite{Clayton} and Oakes \cite{Oakes}). This method has been generalized with two or more frailty variables by Marshall and Olkin \cite{Marsh} and by Joe \cite{Joe}, but their works have not always been  well comprised and often rediscovered. In 1995, Koudraji \cite{Khoud}  developped a procedure   to contruct asymmetric copulas without using a mixing variable, but as a product of two copulas. His work has been generalized by Liebscher \cite{Liebsch} to multivariate copulas.

~

\noindent
Here, our purpose is to show how to construct a copula by using  a priori knowledge on the studied bivariate distribution. 
 We propose a way to construct, step by step, from any basic model more complex models  verifying the  assumption of asymmetry, as well as the assumption of extremal dependance for  minimum and/or  maximum. These models with increasing complexity are obtained using mixing procedures. Each considered assumption adds a new parameter to the model and we can control how this parameter acts: it is not a blind method. This  procedure  is illustrated by using three known copula models which are fit  by maximum likelihood.

~

\noindent
Within the same family the models are nested, so  they can be compared  with likelihood ratio tests. For others comparisons, the best model for a given problem is selected using the Bayesian information criterion (BIC) [\cite{Schwarz}{1978}].
 
~

\noindent
The proposed modelling method is illustrated on buoy data.  In oceanometeorology, scientists are interested in modelling the  statistical dependence between sea state parameters ( e.g. significative height and period of waves, surge, wind speed ...) because it is used in order to study the reliability and fatigue of structures. In this modelling  method, it is important  to take into account the  extreme dependence between the different processes. Namely, the simultaneous occurence of extreme events 
can  be the cause of great environemental or structural damage. Futhermore, the dependence between the different processes is rarely symmetric:  there often  exists an order relation between the variables. For example, one cannot observe
very high waves with very short periods and, on the contrary, far from a storm in time or in space, waves have generally  small height and long period.
However, most models used in oceanography are symmetric and often multivariate Gaussian.
[\cite{Schuel}{2003}]. In 1995, Athanassoulis {\it et al.} proposed bivariate models based on Plackett's copulas.

~

\noindent
In section two, we describe  the  construction of  bivariate  asymmetric distributions with or without extreme dependence and introduce three illustrative copulas that we propose to evaluate. In the third section, we recall how to simulate distributions from  models with mixing variables and assymetric distributions. Such simulation tools are useful for any Monte Carlo approach, for instance. The inference and validation methods are detailed in the fourth section. Finally, in the last section, we described the metocean dataset and present the results of the evaluation of the models.

\section{ Construction of asymmetric distributions with or without extreme dependence} 
To construct any bivariate distribution, copulas allow for the  separate modelling of the univariate margins and the dependence between the variables under weak assumptions. Copulas are a flexible tool for modelling any shape of dependence between two variables, the univariate distributions of these variables being characterized separately.
In the same manner, the estimations of the parameters of the joint model can be made in two steps: the parameters of the univariate margins are estimated firstly and those of the copula secondly.
Here, we show in particular how, using  a priori knowledge on a bivariate distribution -for example, existence of an assymetry in the dependence of the variables, or existence of extreme dependence for the minimum or for the maximum-, we can transform a basic copula into a more complex copula verifying the a priori knowledge.

~

\noindent
Let us first recall some definitions.

\begin{enumerate}
\item  Definition of the Copula \\
The copula  summarizes the  dependence between the two variables.

Following the Sklar theorem (\cite{Sklar}{1959}), to a cumulative distribution function $H(x,y)$ with continuous margins  $F_{1}(x)$ and $F_{2}(y)$, one associates  copula 
$C(u,v)$, defined by 
 \[
 H(x,y)=C(F_{1}(x), F_{2}(y))
 \]
It is easy to verify that the copula is, therefore, a cumulative distribution function (cdf) defined on the square unit with uniform margins. And that it  summarizes the  dependence between the two variables.

When the cdf $H(x,y)$  is derivable and if $(X,Y)$ admits marginal densities  $f_{1}(x)$  and $f_{2}(y)$ with respect to  Lebesgue's measure and a joint probability density function $h(x,y)$, then the theorem of Sklar can be rewritten as 
\begin{equation}
 h(x,y)=f_{1}(x)f_{2}(y)c( F_{1}(x), F_{2}(y))
 \label{eq:jpdf}
 \end{equation}
 where $c(u,v)$ is the density of the copula.
  
  When $C(u,v) \ne C(v,u)$, the copula is said to be \textit{non exchangeable}, which is the situation of assymetry.

~

\item The function of dependence and the measure of extreme dependence \\
Let  $(X_{i}, Y_{i}), \; i=1,...,n $, be a sample of a distribution $H(x,y)$.  To study  extreme events, one considers 
 the distribution of the couple  
 $$\left( \frac{X_{max} -b_{1n}}{a_{1n}}, \frac{Y_{max}-b_{2n}}{a_{2n}}\right)$$
  where $X_{max}= \max(X_{1},...X_{n})$, $Y_{max}= \max(Y_{1},...Y_{n})$. Constants  $a_{in}$ and $b_{in}$, $i=1,2$,
are normalizing constants depending on the margins of  $X$ and $Y$.
One defines then  
$$H_{max}(x,y)=\lim_{n \to \infty} \left(H\left(a_{1n}x+b_{1n}, a_{2n}y+b_{2n}y\right)\right)^{n}$$ 
and its  associated copula $C_{max}(u,v)$. 

 $H(x,y)$ is said to belong to the domain of attraction of $H_{max}(x,y)$.
 If the  distribution $H_{max}(x,y)$ is not the product of the margins, $H(x,y)$ is said to be 
 asymptotically  dependent for the maximum.

 The copula $C_{max}(u,v)$ is such that
 $$C_{max}(u,v) = \lim_{n \to \infty}\left(C\left(u^{\frac{1}{n}},v^{\frac{1}{n}}\right)\right)^{n}$$
 Indeed, if $U_{max}=\max(U_{1},...,U_{n})$ and $V_{max}=\max(V_{1},...,V_{n})$, where the random sample $(U_{1},V_{1}),...(U_{n},V_{n})$ comes  from  copula $C(u,v)$ then the copula associated to $(U_{max},V_{max})$ is $$C^{n}(u^{\frac{1}{n}},v^{\frac{1}{n}})$$

Following Pickands \cite{Pickands}, the associated  copula   $C_{max}(u,v)$ can be written as
$$C_{A}(u,v)=\exp\left(\log uv.A(\frac{\log u}{\log uv})\right)$$
 where $A(.)$ is the dependence function verifying
 $$A: [0,1] \to [\frac{1}{2}, 1], \ \ 
 A \mbox{ is convex, } \max(t,1-t)\leq A(t) \leq 1, \mbox{ and } A(0)=A(1)=1.$$
 The particular case where $A(t) \equiv 1$ corresponds to the 
independence case.

 The  extreme dependence can be quantified by  
 $$\lambda=\lim_{u \to 1}\frac{\bar{C}_{A}(u,u)}{1-u}=2-\lim \frac{\log \bar{C}_{A}(u,u)}{\log u} =2(1-A(\frac{1}{2}))$$ 
 where  $\bar{C}(u,v)$ is the survival
 function of the copula. If a copula  $C(.,.)$ is in the attraction domain of a copula
 $C_{A}(.,.)$, then they have the same  value of $\lambda$ (Joe, page 178). This quantity, when it is strictly positive,  characterizes the asymptotic dependence.
This measure, however, does not seem  adequate for non exchangeable variables, where the maximal dependence is not along the first  diagonal.

If we have to model the extreme dependence for the   minimum of two variables instead of the maximum,   the  dual couple $(1-U,1-V)$  can be  considered in place of
 $(U,V)$. The survival function of the one is the cumulative distribution function of the other. The dual extreme measure is then:
 $$\bar{\lambda}=\lim_{u \to 0}\frac{C(u,u)}{u}.$$

\end{enumerate}

\subsection{The mixture models and its generalizations}

\subsubsection{The frailty model}
To model the dependence between two random variables   $X$ and  $Y$, with cumulative distribution functions (cdf) $F_{1}(x)$ and  $F_{2}(y)$, a usual method
[\cite{Oakes}{1989}] 
is to  suppose  that the two variables are conditionally independent  from a positive ``frailty'' variable $Z$ with cdf $G$:
\[
 H(x,y)=\int F_{1}(x)^{z}F_{2}(y)^{z}dG(z)
 \]
The two margins of  $H(.,.)$ are  $H_{1}(x)=\int F_{1}(x)^{z}dG(z)$ and  $H_{2}(y)=\int F_{2}(y)^{z}dG(z)$.
Calling  $\varphi^{-1}$, the Laplace transform of   $G$, the cdf  $H(.,.)$ can be rewritten as:
 \[
 H(x,y)=\varphi^{-1}( \varphi(H_{1}(x)) + \varphi(H_{2}(y))).
 \]
 The associated copula is then: $$C(u,v)=\varphi^{-1}( \varphi(u) + \varphi(v))$$ 
which is a particular case of an Archimedean copula with generator  $\varphi$  (\cite{Genest2}).

~

\noindent
The frailty models have been defined in the context of lifetime data analysis, from the two survival margins $S_{1}$ and $S_{2}$. In the square unit, this is written as:
\[\bar{C}(u,v)=\varphi^{-1}( \varphi(1-u) + \varphi(1-v)) \]
which is the dual copula of the previous copula.

~

\noindent
Now,  two examples of Archimedean copulas, that we will use in the sequel, are introduced.
\begin{itemize}
\item[(a)]  Clayton's copula \\
If $Z$ has a Gamma distribution with Laplace transform 
$$\varphi^{-1}(t)=(1+t)^{-\frac{1}{\alpha}}, \ \ \alpha \geq 0$$
this results in Clayton's copula:
\begin{equation} 
C_{C}(u,v)=( u^{-\alpha} +v^{-\alpha}-1)^{-\frac{1}{\alpha}}
\end{equation}
This copula owns extreme dependence on the minimum with 
\begin{equation}
\bar{\lambda}=2^{-\frac{1}{\alpha}}
\label{eq:Clambda}
\end{equation}
Written with survival functions, it leads to extreme dependence on the maximum.
The dependence  increases with $\alpha$. When  $\alpha$ tends to $\infty$, the copula tends to the upper maximal dependence copula. The case $\alpha=0$ corresponds to   independence. Kendall's tau is equal to $\frac{\alpha}{\alpha+2}$.

\item[(b)] Gumbel's copula \\
If $Z$ has a positive stable distribution   
$$ \varphi^{-1}(t)=\exp\left(-t^{\alpha}\right), \ \ 0<\alpha \leq 1$$
one obtains   Gumbel's copula, 
 \begin{equation}
 C_{G}(u,v)= \exp \left[ -((-logu)^{\frac{1}{\alpha}}+ (-logv)^{\frac{1}{\alpha}}))^{\alpha}\right]
\end{equation}
which is the only Archimedean and extreme value copula with 
\begin{equation}
\lambda=2-2^{\alpha}
\label{eq:Glambda}
\end{equation}
 Its lower tail dependence is zero. Its Kendall's tau is given by $\tau=1-\alpha$. 
The dependence  decreases according to $\alpha$.
\end{itemize}

\subsubsection{ Joe's generalization}

A max-infinitely divisible (max-id) bivariate cdf $F$  is such that any power of it, $F^{\gamma},\;\gamma>0$, is still a cdf.
Joe  [\cite{Joe}{1997}] generalizes the mixture model with any  max-id copula $K(u,v)$ in place of the product of the marginals, and with $\varphi^{-1}$, the Laplace transform of a frailty variable $Z$. The  obtained copula $\hat{C}(u,v)$ verifies:

\[
\hat{C}(u,v)=\int K^{z}dG(z)=\varphi^{-1}\left(-\log K(e^{-\varphi(u)},e^{-\varphi(v)})\right)
\]

\subsubsection{ Marshall and Olkin procedure}
Marshall and Olkin \cite{Marsh} have proposed another generalization of this method  using two frailty variables $Z_{1}$ and $Z_{2}$. Specifically, let    $G(.,.)$ be a cdf
such that   
$\bar{G}(0,0)=1$ with margins  $G_{i}(.)$, $i=1,2$. Then  define new copula $C(.,.)$ by 
$$C(u,v)=\int\int (F_{1}(u))^{z_{1}}(F_{2}(v))^{z_{2}}dG(z_{1},z_{2}) $$
where  $F_{1}(u)=\exp(-\varphi_{1}(u))$ and  $F_{2}(v)=\exp(-\varphi_{2}(v))$ with $\varphi_{i}^{-1}$, $i=1,2$, the Laplace transforms of   $G_{i}$.

~

\noindent
  They presented examples with  frailty variables  $Z_{1}$ and $Z_{2}$ such that
$$Z_{1}=U_{1}+W \mbox{ and } Z_{2}=U_{2}+W$$
where  $W$, $U_{1}$, $U_{2}$ are independent random variables. Let  $\psi_{1}$, $\psi_{2}$ and $\psi_{0}$, be the Laplace transforms of the three variables $U_{1}$, $U_{2}$ and $W$. Then, the copula is written 
$$ C(u,v)=\psi_{1}( \varphi_{1}(u)) \ \psi_{2}(\varphi_{2}(v)) \ \psi_{0}\left( \varphi_{1}(u)) + \varphi_{2}(v)\right)$$
with $\varphi_{i}^{-1}(t)=\psi_{i}(t)\psi_{0}(t)$, $i=1,2$.

~

\noindent
Let us now present two examples where parameters  $\alpha_{0}$, $\alpha_{1}$ and $\alpha_{2}$ are associated to the three  Laplace transforms $\psi_{i}$,  $i=0,1,2$:

\begin{itemize}
\item[(a)] Clayton's family  extension  \\
Let $W$, $U_{1}$ and $U_{2}$ be  three variables with Gamma Laplace transforms with parameters $\alpha_{0}$, $\alpha_{1}$ and $\alpha_{2}$ and let  $Z_{i}=U_{i}+W$, $i=1,2$.  The Laplace transform of  $Z_{i}=U_{i}+W$ is given by
 $$\varphi^{-1}_{i}(t)=(1+t)^\frac{-1}{\alpha_{0}+\alpha_{i}} \mbox{ for } i=0,1,2$$
  and 
\[\tilde{C}(u,v)=u^{\frac{\alpha_{0}}{\alpha_{1}+\alpha_{0}}} v^{\frac{\alpha_{
0}}{\alpha_{2}+\alpha_{0}}}(u^{-\frac{\alpha_{0}\alpha_{1}}{\alpha_{0}+\alpha_{1}}} + v^{-\frac{\alpha_{0}\alpha_{2}}{\alpha_{0}+\alpha_{2}}}-1)^{-\frac{1}{\alpha_{0}} } \]

Using the reparametrization $\theta=\frac{\alpha_{1}}{\alpha_{0}+\alpha_{1}}$ and $\delta=\frac{\alpha_{2}}{\alpha_{0}+\alpha_{2}}$, the copula can be rewritten as:
\begin{equation}
 \tilde{C}(u,v)=u^{1-\theta}v^{1-\delta}C_{C}(u^{\theta},v^{\delta})
 \label{eq:clayton}
\end{equation}
where $C_{C}(u,v)$ is the Clayton's copula.

~

\item[(b)] Gumbel's family extension \\
Considering two frailty variables and Laplace transforms: $\psi_{i}(t)=\exp(-\theta_{i}s^{\alpha})$  with $0<\alpha \leq 1$ and $\theta_{i} \geq 0$ for $i=0,1,2$, one obtains:
\begin{eqnarray}
C(u,v)&=&\exp\left( \frac{\theta_{1}\log u}{\theta_{1}+\theta_{0}} + \frac{\theta_{2}\log v}{\theta_{2} +\theta_{0}} - \theta_{0}[(-\frac{\log u}{\theta_{1}+\theta_{0}})^{\frac{1}{\alpha}} + (-\frac{\log v}{\theta_{2}+\theta_{0}})^{\frac{1}{\alpha}}]^{\alpha} \right) \nonumber  \\
      &=&
      u^{\frac{\theta_{1}}{\theta_{1}+\theta_{0}}}v^{\frac{\theta_{2}}{\theta_{2}+\theta_{0}}}\exp\left(-\theta_{0}[(-\frac{\log u}{\theta_{1}+\theta_{0}})^{\frac{1}{\alpha}} + (-\frac{\log v}{\theta_{2}+\theta_{0}})^{\frac{1}{\alpha}}]^{\alpha} \right) \nonumber  
\end{eqnarray}
Using the reparametrization, $\theta_{0}=1$, $\theta=\frac{1}{1+\theta_{1}}$ and  $\delta=\frac{1}{1+\theta_{2}}$, one obtains the same formal writing as in   equation (\ref{eq:clayton}), and the obtained copula corresponds to   Tawn's  bivariate extreme value  distribution (assymetric bilogistic distribution)  [\cite{Tawn}] 
\begin{equation}
 \tilde{C}(u,v)=u^{1-\theta}v^{1-\delta}C_{G}(u^{\theta},v^{\delta})
 \label{eq:gumbel}
\end{equation}
where $C_{G}(u,v)$ is  Gumbel's copula.
\end{itemize}

\subsection{Constructing asymmetric dependence}
The procedure of Marshall and Olkin (\cite{Marsh}) and its extension can be replaced in the framework of the
 assymetrization procedure  proposed by Khoudraji (in his thesis [\cite{Khoud}{1995}]).
 Without using frailty variables,  he   constructs an asymmetrized copula from two  symmetric copulas.
 
 ~
 
 \noindent
If  $C_{1}(u,v) $  and  $C_{2}(u,v) $ are two   symmetric copulas, let $\tilde{C}(u,v)$ be:
\[\tilde{C}(u,v)=C_{1}(u^{1-\theta}v^{1-\delta})C_{2}(u^{\theta},v^{\delta}),\; 0 \le \theta,\, \delta  \le 1 \]
One sees easily that, except for particular cases,  $\tilde{C}(u,v) \ne \tilde{C}(v,u)$.

~

\noindent
A particularly interesting case is when  $C_{1}(u,v)$ is the independence copula:
\begin{equation} 
 \tilde{C}(u,v)=u^{1-\theta}v^{1-\delta}C_{2}(u^{\theta}v^{\delta})
 \label{eq:general}
\end{equation}
 In this last case, the method weakens the dependence, especially if $C_{2}(u,v)$  is an extreme value copula for the maximum, then the new copula $\tilde{C}(u,v)$
is still an extreme value copula, but the parameter of extreme dependence  $\tilde{\lambda}$ is smaller than $\lambda$ [\cite{Drouet}{2004}]. This phenomena occurs for instance in the case of Gumbel's copula.

~

\noindent
Using  Fréchet's bounds, we have
$$\tilde{C}(u,v)=u^{1-\theta}v^{1-\delta}C_{2}(u^{\theta},v^{\delta}) \leq u^{1-\theta}v^{1-\delta}min(u^{\theta},v^{\delta})$$
The right hand side member of this inequality is the Cuadras-Augé copula (see \cite{Khoud}) whose Kendall $\tau$ is equal to  
\begin{equation}
\frac{\theta\delta}{\theta+\delta-\theta\delta}
\label{eq:cuad}
\end{equation} 
Consequently  $\tau_{\tilde{C}}$ is smaller than this last quantity.
If the symmetric copula has no extreme maximal  dependence, then  the assymmetrized copula does.
Moreover, even if $C_{2}(u,v)$ has a lower tail dependence, the lower tail dependence of $\tilde{C}(u,v)$ is zero.

~

\noindent
In a few examples, we can suppose that a common cause acts on the two variables $U$ and $V$, but that one variable has its proper variability cause. 
In such cases, one can write $Z_{1}=U_{1}$ and $Z_{2}=U_{2}+W$ and one obtains the simpler model 
[\cite{Drouet}{2004}]:
\begin{equation}
\tilde{C}(u,v)=v^{1-\delta}C(u,v^{\delta}),\; 0 \le \delta \le 1. 
\label{eq:2par}
\end{equation}
When $\delta=1$, we retrieve the basic copula. We are, therefore, able to deduce a test for the asymmetry.

~

\noindent 
Using equation (\ref{eq:general}) or (\ref{eq:2par}),  we can construct three assymetrized copulas on the basis of three basic copulas:

\begin{enumerate}

 \item  Plackett's copula $C_{p}(u,v)$.\\
\begin{equation}
C_{p}(u,v)=\frac{1}{2\alpha}\left(1+\alpha(u+v)-[(1+\alpha(u+v)^{2}-4\alpha(\alpha+1)uv]^{\frac{1}{2}} \right),\; -1 \le \alpha
\end{equation}
The Plackett copula is used, for instance, in oceanometeorology to model the dependence between  the couples of wave heights and wave periods \cite{Athan}. This copula is not obtained from a frailty model and has neither dependence for the maximum nor for the minimum. It is introduced here for comparison. 

\item Claton's survival copula\\
If we apply the former procedures to the survival  Clayton's copula 
\begin{equation}
S_{c}(1-u,1-v)=\left( (1-u)^{-\alpha} +(1-v)^{-\alpha}-1\right)^{-\frac{1}{\alpha}},\; \alpha \geq 0
 \end{equation}
we obtain 
\begin{equation}
\tilde{S}(1-u,1-v)=(1-u)^{1-\theta}v^{1-\delta}S_{c}((1-u)^{\theta},(1-v)^{\delta})
\label{eq:clayta2}
 \end{equation}
 
 or 
 \begin{equation}
\tilde{S}(1-u,1-v)=(1-u)^{1-\theta}S_{c}((1-u)^{\theta},(1-v)^{\delta})
\label{eq:clayta}
 \end{equation}
 
As Clayton's model is constructed on survival copula, the assymetry is applied to the  variable $(1-u)$, and not to $v$.
The resulting copula shows no  extreme dependence  for the maximum [\cite{Drouet}{2004}].

\item Gumbel's copula \\
The third copula is Gumbel's copula, which allows us to test for extremal dependence for the maximum. See earlier (\ref{eq:gumbel}) or 
\begin{equation}
 \tilde{C}(u,v)=v^{1-\delta}C_{G}(u^{\theta},v^{\delta})
\end{equation}
\end{enumerate}

\subsection{Extreme dependence for the minimum}

The three  models under consideration do not include any extreme dependence for the minimum. Using  Joe's  mixture  method, we can construct new models with  minimal dependence. 

~

\noindent
Introducing  $Z$, a mixture variable with Gamma Laplace transform $\varphi^{-1}$ whose parameter is $\beta$  and from $\tilde{C}(u,v)$, the asymmetrized copula   we write 
\begin{equation}
 \hat{C}(u,v)=\varphi^{-1}\left(-\log \tilde{C}(e^{-\varphi(u)},e^{-\varphi(v)})\right) \\
            =(1-\log\tilde{C}(e^{-\varphi(u)},e^{-\varphi(v)}))^{-\frac{1}{\beta}}, \; \beta \geq 0
            \label{eq:hatC}
 \end{equation}
 If  parameter $\beta=0$, we retrieve the model without minimum dependence. We can then deduce a test for this dependence. In the other cases, this copula has a lower tail dependence greater than the lower tail dependence of the Clayton copula (see Appendix).

~

\noindent
For the Gumbel copula, we obtain:
\begin{equation}
 \hat{C}(u,v)=\varphi^{-1}\left((1-\theta)\varphi(u) + (1-\delta)\varphi(v) + \{(\theta \varphi(u))^{\frac{1}{\alpha}} +(\delta \varphi(v))^{\frac{1}{\alpha}} \}^{\alpha} \right) 
 \end{equation}
Its lower tail dependence is given by
\[\bar{\lambda}_{\hat{C}}=r^{-\beta}\]
with $r=1-\theta+(\theta^{\alpha}+1)^{\frac{1}{\alpha}}$ (see Appendix)

~

\noindent
In case of any  bivariate extreme value copula, $C_{A}(u,v)=\exp ( \log(uv)A(\frac{\log u}{\log uv}))$, we obtain:
\[\tilde{C}(u,v)=u^{1-\theta}v^{1-\delta}C_{A}(u^{\theta},v^{\delta}) \]
and then:
\[ \hat{C}(u,v)=\varphi^{-1}\{ (1-\theta)\varphi(u) + (1-\delta)\varphi(v) + \{\theta \varphi(u)+\delta \varphi(v)\}A (\frac{\theta \varphi(u)}{\theta \varphi(u)+\delta \varphi(v)} )\} \]

~

\noindent
When a copula, such as Clayton's copula is assymetrized  from its survival function, 
we use   $ \tilde{C}(u,v)=-1+u+v+ \tilde{S}(u,v)$  to construct:

\[ \hat{C}(u,v)=\varphi^{-1}\{-\log \tilde{C}( e^{-\varphi(u)},  e^{-\varphi(v)}) \} \]

\section {Simulations}

\subsection{Method}
The articifial random generation of samples following the proposed distributions  may be useful for Monte Carlo testing.  
Many simulation methods have been developped in many particular  cases:  Archimedean copulas (see for example the papers of Genest and Mackay \cite{Genest2}, Genest and Rivest\cite{Genest3}), extreme value distributions (Ghoudi et al. \cite{Ghoudi}, Shi \cite{Shi}, Stephenson \cite{Stephens}),  mixtures of distributions (Marshall and Olkin \cite{Marsh}). In other cases, a general procedure can be used.

~

\noindent
  Different simulation methods will now be detailed.
\begin{enumerate}
\item  Mixtures of distributions 

 Here, we work with a  frailty variable following a  Gamma distribution with parameter $\alpha$, but another distribution could be used.

This is the case of Clayton's copula (case $1$), and when we add an extreme dependence for the minimum to  Gumbel's assymetrized distribution (section 2.2, case $2$).
We use in that case the procedure of Marshall and Olkin \cite{Marsh} for mixture distributions, as follows.

\begin{enumerate}
\item Simulate a random sample of couples $(U_{i},V_{i})$ of independent variables or according to the distribution $\tilde{C}$.

\item Generate a random sample of the mixture Gamma variable $Z_{i}$ with parameter $\alpha$.

\item Construct $S_{i}=((1-\frac{1}{Z_{i}})\log(U_{i}))^{\alpha}$ and $T_{i}=((1-\frac{1}{Z_{i}})\log(V_{i}))^{\alpha}$. $S_{i}$ and $T_{i}$ have  Clayton's distribution or the $\hat{C}$ distribution.

\end{enumerate}

\item Bivariate Extreme Value Distribution (case of Gumbel's copula)

 It is possible to generate a sample according to Gumbel's copula  using a frailty variable with positive stable distribution  Ps($\alpha$) with parameter $\alpha$. See for example A. Stephenson \cite{Stephens} for generation  Ps($\alpha$) distributions. 
  
  ~
  
  \noindent
  Instead of that, we have chosen here a procedure derived from Lee \cite{Lee} for generating logistic extreme value distributions.
 The procedure uses the fact that $T=\frac{\varphi(U)}{\varphi(U)+\varphi(V)}$, with $\varphi(U)=(-log(U))^{\alpha}$, has a uniform distribution  and $T$ and $Z=C(U,V)$ are independent. Furthermore $Z_{1}=(\varphi(Z))^{\alpha}$ is distributed as a mixture of two Gamma variables.

  ~
  
  \noindent
  This method can be generalized to more than two variables (trilogistic distributions,...) \cite{Stephens} and also to extreme value distributions different from logistic distributions (\cite{Ghoudi}) using the two variables $T=\frac{log(U)}{log(U)+log(V)}$ and $Z=C(U,V)$ which are not independent but whose joint distribution is a function of the dependence function $A(.)$ (see section 2) and of its second order derivative $A''(.)$.

\begin{enumerate}
\item Simulate a mixture of  Gamma variables $\Gamma_{i}$ with parameter $(1,1)$ and $(2,1)$. The $\Gamma(2,1)$ is generated in the proportion $\alpha$. And let $Z_{i}=\Gamma_{i}^{\alpha}$.
\item Simulate a random sample of uniform variables $W_{i}$ and construct the products $J_{i}=W_{i}Z_{i}$.
\item Let  $U_{i}=\exp(-J_{i}^{\frac{1}{\alpha}})$ and $V_{i}=\exp(-(Z_{i}(1-W_{i})^{\frac{1}{\alpha}})$. 
 $U_{i}$ and $V_{i}$ have the Gumbel distribution       
\end{enumerate}
         
\item Assymmetrization from the cdf $C_{1}$ and $C_{2}$ with exponent $\theta$ and $\delta$.

This procedure is described in Khoudraji \cite{Khoud}. The idea is that if $(U_{1},V_{1})$ and $(U_{2},V_{2})$ have respectively  cdfs $C_{1}$ and $C_{2}$, then $\max(U_{1}^{\frac{1}{1-\theta}},U_{2}^{\frac{1}{\theta}})$ and 
$\max(V_{1}^{\frac{1}{1-\delta}},V_{2}^{\frac{1}{\delta}})$ have the cdf $\tilde{C}$.

 \begin{enumerate}
 \item Simulate a random sample of couples $(U_{i},V_{i})$ according to the symmetric distribution $C_{1}$
 
 \item Calculate $W_{i}=U_{i}^{\frac{1}{1-\theta}}$ and $X_{i}=V_{i}^{\frac{1}{1-\delta}}$

 \item Simulate a random sample of couples $(S_{i},T_{i})$  from the symmetric distribution $C_{2}$.

 \item Calculate $Y_{i}=S_{i}^{\frac{1}{\theta}}$ and $Z_{i}=T_{i}^{\frac{1}{\delta}}$

 \item Choose $\tilde{U}_{i}=\max(W_{i},Y_{i})$ and $\tilde{V}_{i}=\max(X_{i},Y_{i})$

 $(\tilde{U}_{i},\tilde{V}_{i})$ have the distribution $\tilde{C}$.
 \end{enumerate}
 
 \item  General Procedure
 
 The simulation of couples $(U_{i},V_{i})$ from  $\hat{C}(u,v)$ (eq. 11) is obtained by a more general method.
 \begin{enumerate}
 
 \item  Simulate a random sample of couples $(U_{i},T_{i})$ of independent uniform variables.
     
 \item Let $\hat{C}_{2|1}(t|u)=\frac{\partial{C(u,t)}}{\partial{u}}$, the conditional distribution of $\hat{C}(u,v)$. Make the transformation $ V_{i}=\hat{C}_{2|1}^{-1}(T_{i})$. Then $(U_{i}, V_{i})$ are sampled from $\hat{C}(u,v)$. When no simple analytical expression is available for $\hat{C}_{2|1}^{-1}$, then a numerical solution of $v_{i}=\hat{C}_{2|1}(u_{i},t_{i})$ is looked for. This is the case when we use equation (\ref{eq:hatC}) from Clayton's and Plackett's assymetrized models.
    \end{enumerate}

\end{enumerate}
 \subsection{Illustration}
 
\begin{figure}
  \centering
  \includegraphics[scale=.5]{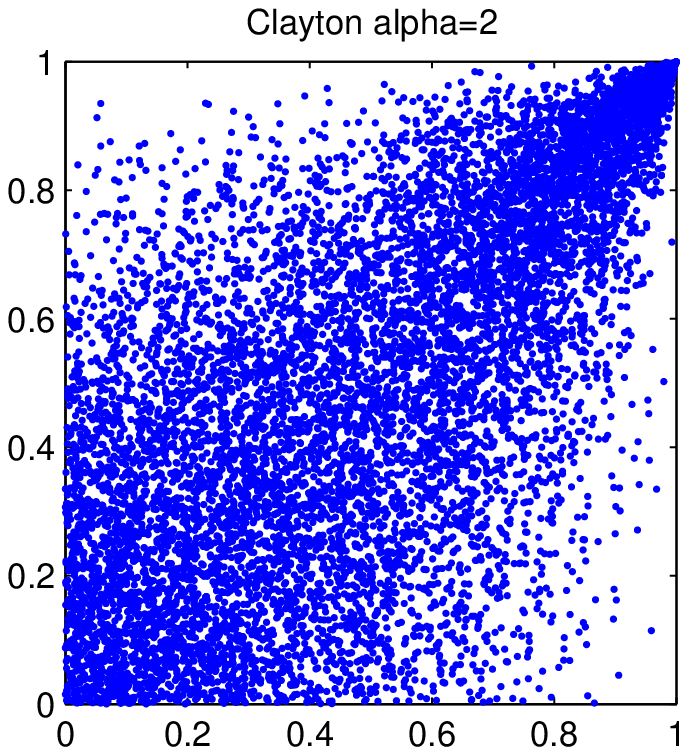}
  \includegraphics[scale=.5]{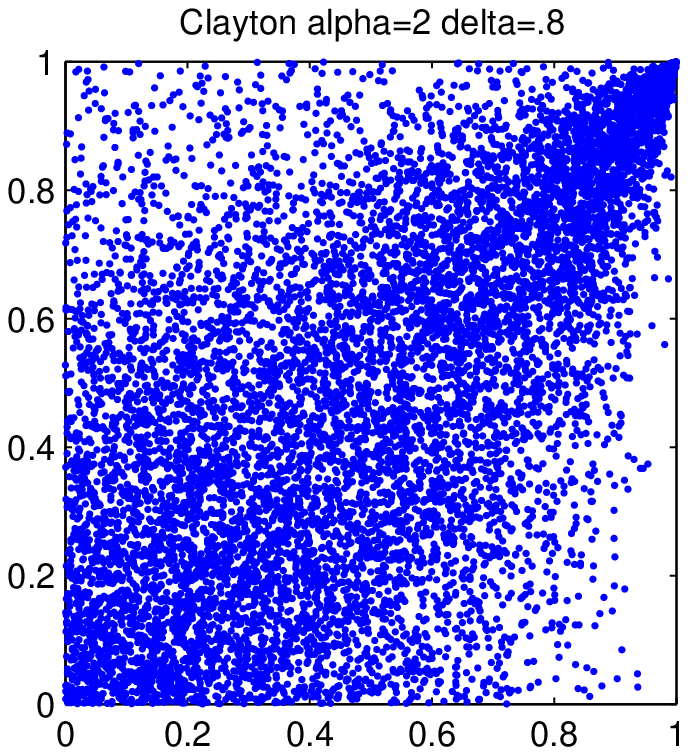}  
  \includegraphics[scale=.5]{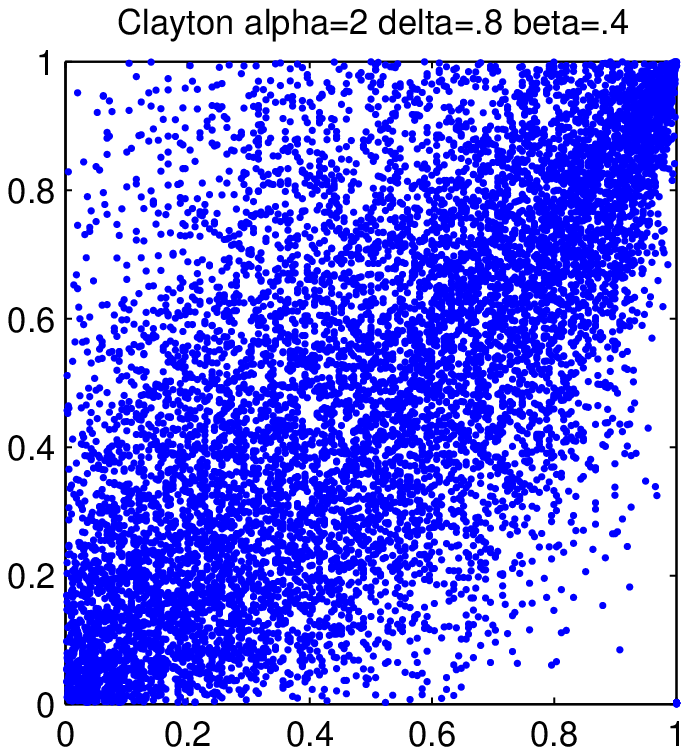}
   \caption{\label{fig:claytonsim} {\it Three datasets simulated from the Clayton family with one parameter (left), two parameters (middle) and three parameters (right). One has respectively $\tau_{1}=.50$, $\tau_{2}=.44$, $\tau_{3}=.50$. }}
\end{figure}

  In figure \ref{fig:claytonsim}, we present three examples of generated datasets from  Clayton's survival  copula, with one, two or three parameters.  Parameter $\alpha$ is the same for the three datasets.
  Assymetry parameter $\delta$ (the other parameter $\theta$ is fixed to one) is the same for the second and the third datasets. We can then see how each parameter acts  on the dependence.
 The dataset with one parameter  is sampled from a distribution with  Kendall's tau equal  to $\tau=0.50$.
 The dataset  generated with two parameters $\delta$ and $\alpha$ has its Kendall tau  bounded by the assymetry parameter $\delta$ (see eq.\ref{eq:cuad}). Here, it is equal to $0.44$.
 When we add  parameter $\beta$ corresponding to minimal dependence, the dependence on the third dataset becomes larger than in the second case and $\tau$ becomes equal to 0.50.
 
\section{Inference and comparison of the models}
\label{sec:inference}

The estimation of  parameters is done in two steps: at first estimating the margins by a moment method, then  estimating  the parameters of the copula  by ML. This method, called IFM (Inference For Margins) was developped by Joe \cite{Joe2} and has mostly good properties (consistancy, asymptotic normality for the parameters).
Let $\bf{\theta_{1}}$ and $\bf{\theta_{2}}$, be the parameters of the margins and $\bf{\eta}$, the parameters of the copula, then the loglikelihood $L$ of the sample $(X_{i},Y_{i}),i=1,...n$ can be written using  equation (\ref{eq:jpdf}) as:
\[ L({\theta},{\eta})=\sum_{i=1}^{n}\left\{log(f_{\theta_{1}}(x_{i}))+log(f_{\theta_{2}}(y_{i}))+log(c_{\eta}(F_{\theta_{1}}(x_{i}),F_{\theta_{2}}(y_{i})))\right\}
\]
Generally, there is no closed-form solution to the problem of maximisation of $L$, namely all the parameters are linked by the copula, but the loglikelihood being the sum  of the two terms: the loglikelihood of the margins, and the loglikelihood of the copula,  the maximisation  may be split in two sub problems. In practice, the estimation of the margin parameters is firstly performed  and the obtained estimations $\hat{\theta_{1}}$ and $ \hat{\theta_{2}}$ are subtituted in the last term of the loglikelihood for the estimation of  parameters $\eta$.

 ~
 
\noindent
 Moreover, to avoid misspecifications for the margins and the propagation of this error to the copula, we use a semi-parametric approach to model the margins, already used by Coles and Tawn [\cite{Coles} {1994}].  The idea consists in modelling the distribution of the data over a well chosen high threshold by a generalized Pareto distribution (see Davison and Smith [\cite{Davis} {1990}]) and under the threshold by the empirical distribution.   The generalized Pareto distribution is given by
\[ F_{X}(x)=1-(1-u_{0})[1-\frac{k}{\sigma}(x-x_{0})]_{+}^{\frac{1}{k}},\;x \ge x_{0} \]
where $x_{0}$ is the chosen high threshold  and   $u_{0}$ the corresponding percentile. The main advantage of this model is that it  provides a more general and more realistic copula (the maxima of which being not one) than a fully non parametric model. 
 
 ~
 
\noindent
At the second step, the parameters of the copula are estimated by Maximum Likelihood.  All the ML estimations are obtained by using numerical optimization. The BIC-criterion allows us to compare all the models. Futhermore, inside the same family, the models are nested so if we compare the one parameter and the two parameter model, we can test the hypothesis that  the assymetry is present (i.e. that the assymetry  parameter is equal to one).  We can also test if the minimal dependence is present, testing the $\beta$ parameter at zero.
 
 ~
 
\noindent 
Finally, from the estimation of  $\alpha$ parameter of  Clayton's or  Gumbel's one parameter models, we can deduce an estimation of the upper tail dependence  using formulas (\ref{eq:Clambda}) and (\ref{eq:Glambda}) given in  section 2.1.1  and an estimation  of its variance by the delta-method.
 In the case of  Clayton's copula, the upper tail dependence is estimated as $2^{-\frac{1}{\hat{\alpha}}}$ and its variance is given by $(\frac{1}{\hat{\alpha^{2}}}\ln(2)2^{-\frac{1}{\hat{\alpha}}})^2 \mbox{var}(\hat{\alpha})$.
 
\section{Application}

In order to study the reliability and/or the fatigue of marine structures, engineers need to know the joint distribution of sea state and atmospheric parameters. The sea state represents the state of the marine environnement at a given location and time. It is  described by synthetic parameters like the significant wave height denoted $H_s$ and the mean wave period denoted $T_p$.  It is also  usual  to consider the wind speed $W_s$. For reliability, it is determinant to well model the extremal dependence. For fatigue, the distribution close to the mode is generally of greater importance \cite{Rychlik}. 

~

\noindent
  In order to model the joint distributions, we have selected three candidate models.
 \begin{itemize}
 \item Gumbel's model naturally characterizes a maximal extreme dependence.
 \item Plackett's model was already used to characterize the dependence between significant wave height and mean period of a sea state \cite{Athan}; there is no extreme dependence in this model.
 \item Clayton's model plays the same role as Pareto distribution in univariate case: indeed it is a limit conditional model in the family of Archimedean copula [\cite{Juri}{2002}]. 
 \end{itemize}
 The models have one to four parameters: the basic models have one global dependence parameter, the asymmetrization adds one or two parameters and a last parameter corresponds to the minimal dependence obtained by introducing a mixture Gamma variable. 
 
 ~
 
 In this paper, we do not consider any distribution for more than two variables for two main reasons: engineers mostly use only univariate or bivariate distributions.  However the presented theoretical results could be generalized to trivariate models, after some calculus. 

\subsection{Data description}

In this paper, we consider data of the K1 buoy which is located in the North Atlantic close to the French coast, at the geographic coordinates (48.00N,12.40W). Five years of hourly data are recorded for the three variables: $H_s$, $T_p$, $W_s$ from 2002 to 2007. The $T_{p}$ recording process leads to integer values for this variable. In order to allow for better estimation of the parameters of the generalised Pareto distribution, a uniform noise defined on $[-/2,+1/2]$ has been added to the obeserved $T_{p}$. The transformed $T_{p}$ is also used in the sequel.

\subsection{Model}
The margins are modeled as discussed in section \ref{sec:inference}. The thresholds for the semiparametric transformation chosen empirically so that the Pareto Generalized Distribution good fits the data. In practice, we use the 90\% quantile for $H_s$ and $T_p$ and the 96\% quantile for $W_s$. The parameters estimated by a moment method are reported in table \ref{tab:margin_par}. This estimator has been chosen rather than others because of its robustness in this application.

\begin{table}[h]
\centering
\begin{tabular}{lccc}
\hline
& Location & Scale & Shape \\
\hline
\hline
$H_s$ & 6.10 &    1.07 $\mathit{(0.002)}$ &  -0.07 $\mathit{(0.0009)}$ \\
$W_s$ &  14.90 & 0.92 $\mathit{(0.004)}$ &  -0.11 $\mathit{(0.003)}$ \\
$T_p$ & 9.81   & 0.91 $\mathit{(3e^{-2})}$ & 0.11 $\mathit{(0.02)}$ \\
\hline 
\hline
\end{tabular}
\caption{\label{tab:margin_par} Estimated parameters for the marginal GPD}
\end{table}\begin{figure}[h]
\centering
\includegraphics[scale=.6]{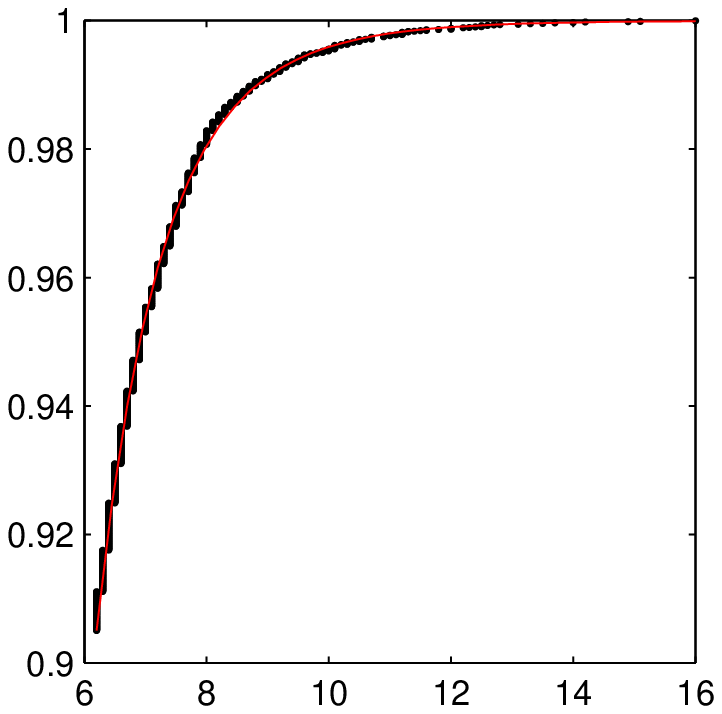}
\includegraphics[scale=.6]{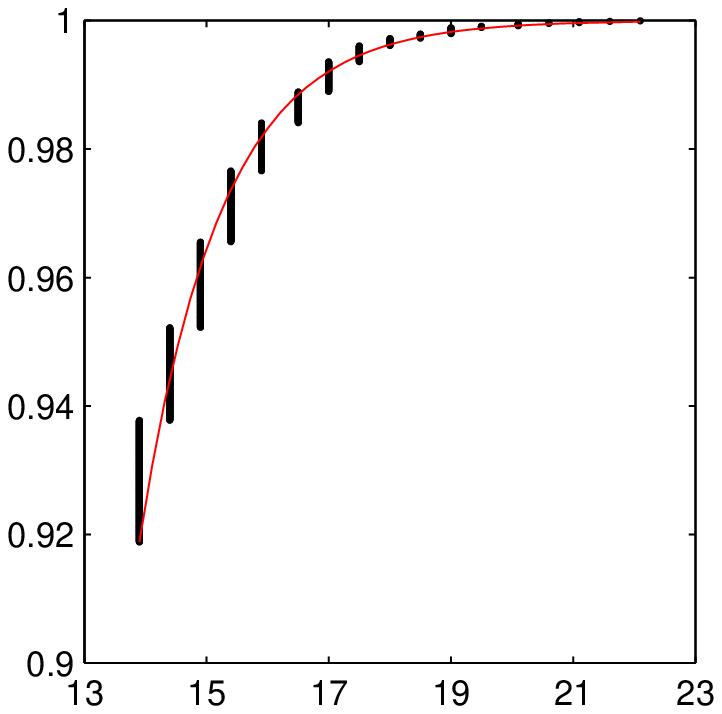}
\includegraphics[scale=.6]{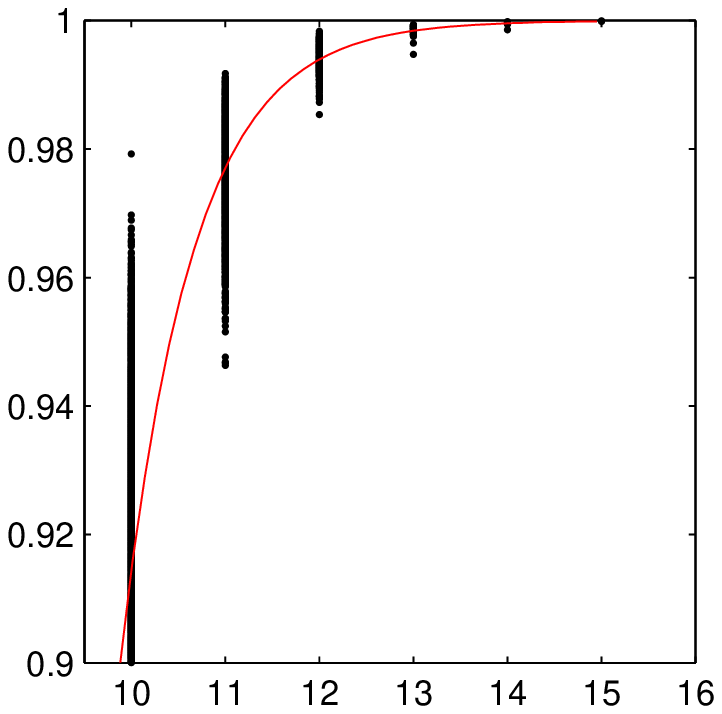}
\caption{\label{fig:margins} Empirical cdf \textit{(points)} and fitted GPD \textit{(line)} for $H_s$ (left), $W_s$ (middle) and $T_p$ (right)} 
\end{figure}

~

\noindent
 The scale parameters are difficult to interpret. The shape parameters of $H_s$ and $W_s$ are as low as expected for these variables. Those of $T_p$ is positive and it represents an extreme distribution of Weibull type.

~

\noindent
Figure \ref{fig:margins} illutrates the fitting of the GPDs on the empirical cumulative distribution functions. The agreement is good.

~

\begin{figure}
  \centering
  \includegraphics[scale=.6]{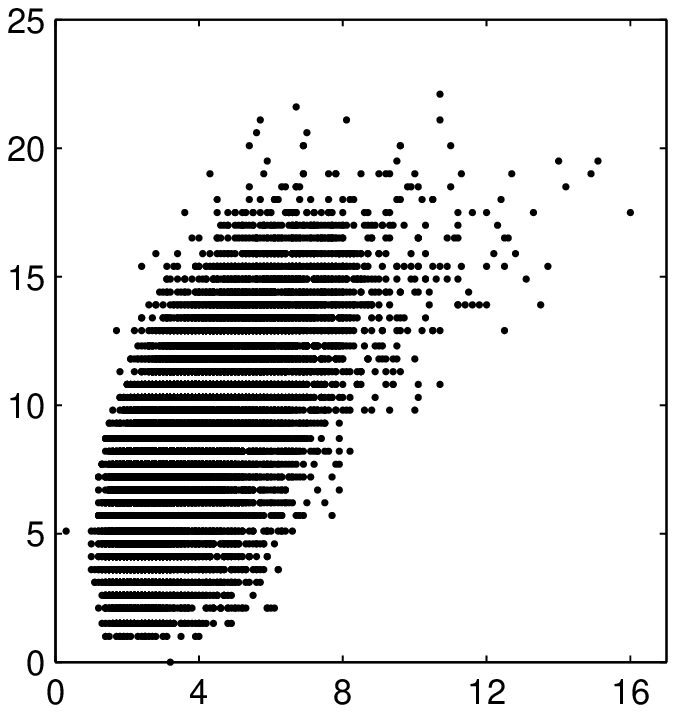}
  \includegraphics[scale=.6]{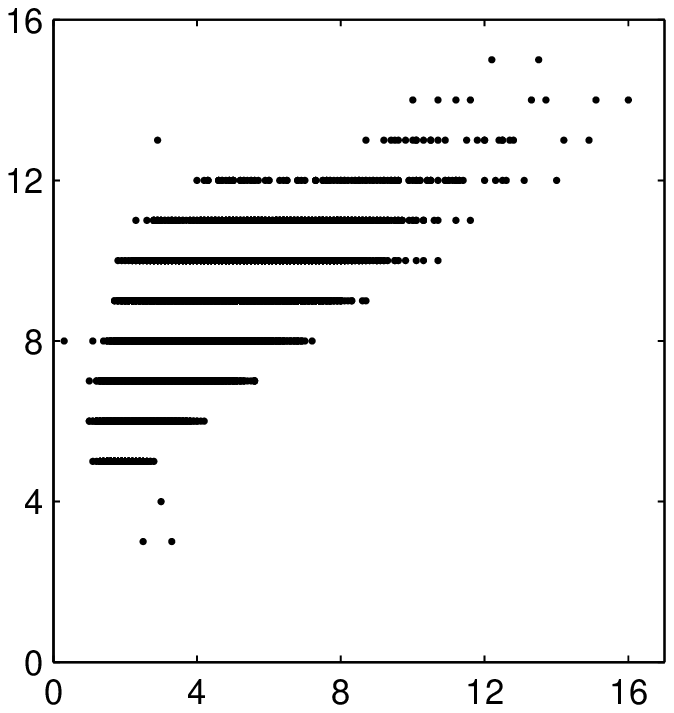} \\  
  \includegraphics[scale=.6]{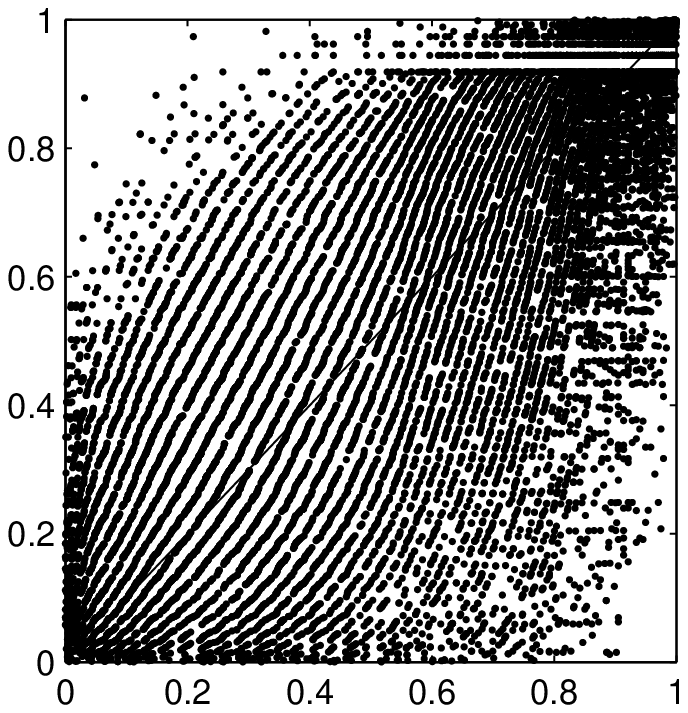}
  \includegraphics[scale=.6]{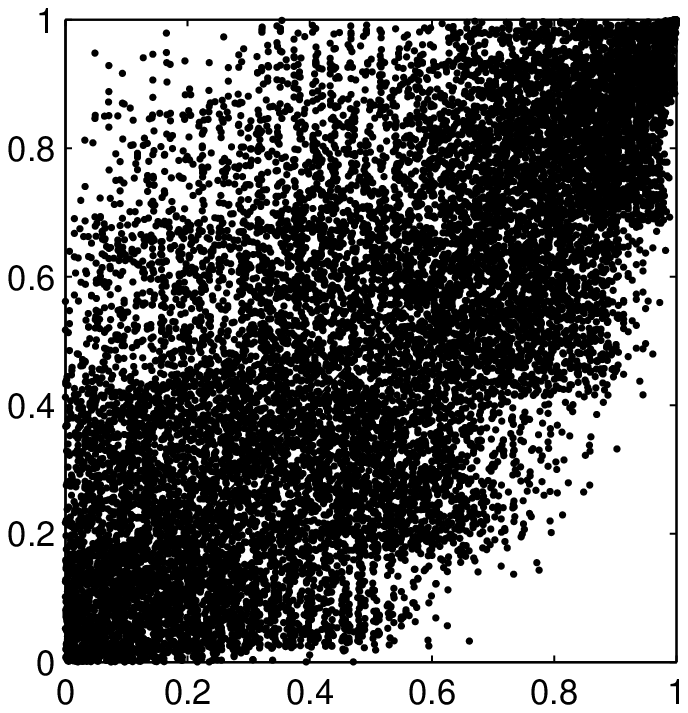}  
  \caption{\label{fig:copulas} Observed distributions (top pannel) and copulas (bottom pannel) of couple $(H_s,W_s)$ (left) and $(H_s,T_p)$ (right)}
\end{figure}

\noindent
The joint distributions and the copulas of pairs $(H_s,W_s)$ and $(H_s,T_p)$ are represented in figure \ref{fig:copulas}. One observes that dependence of the variable $H_s$ with $W_s$ and $T_p$ is quite strong. Both copulas  seem to present extremal dependence and the $(H_s,T_p)$ copula clearly shows asymmetry. The Kendall tau have been estimated and they are respectively equal to 0.47 and 0.52 for $(H_s,W_s)$ and $(H_s,T_p)$ which can be considered as a strong dependence.

~

\noindent
The parameters of the $3 \times 3$ models are estimated as described in section 4 and the results are reported  in tables \ref{tab:par1} and \ref{tab:par2}.
The standard errors are calculated from the Hessian matrix of the log-likelihood. And as it was already remarked, the models are nested so that, when a model is degenerated, the variances of the former model are reported.  In the tables, we also report the value of the log-likelihood at the estimated parameters and the BIC. The log-likelihoods allow for the  comparison of the nested models by  log-likelihood ratio tests. 

%

\begin{table}[!ht]
\begin{center}
\begin{tabular}{cccc }
\hline
\hline
Model & Plackett & Gumbel & Clayton \\
\hline
One parameter $\alpha$  & 6.76 {\it (0.15) } &  0.57 {\it (3.6e$^{-3}$)} & 1.24 {\it (1.7e$^{-2}$)} \\
\textbf{$\log\mathcal{L}$}  & \textbf{4054} & \textbf{4297} & \textbf{4019} \\
\textbf{BIC} &  \textbf{-8099} &  \textbf{-8585} & \textbf{-8029} \\
\hline
Three parameters$\begin{array}{c} \theta \\ \delta \\ \alpha \end{array}$ 
& $\begin{array}{c}  1.00 \\ 0.73 \\ 12.21 \end{array}$  $\begin{array}{c} \mathit{(0.01)} \\ \mathit{(0.02)} \\ \mathit{(0.64)}\end{array}$ 
& $\begin{array}{c} 1.00 \\0.94\\ 0.55 \end{array}$ $\begin{array}{c} \mathit{(1.11)} \\ \mathit{(0.47)} \\ \mathit{(0.23)} \end{array}$ 
& $\begin{array}{c} 0.78\\ 0.96 \\ 2.34 \end{array}$ $\begin{array}{c} \mathit{(0.01)}\\ \mathit{(0.06)} \\ \mathit{(0.07)} \end{array}$ 
\\
\textbf{$\log\mathcal{L}$}  & \textbf{4150} & \textbf{4302} & \textbf{4246} \\
\textbf{BIC} &  \textbf{-8270} &  \textbf{-8577} & \textbf{-8463} \\

\hline
Four  parameters $\begin{array}{c} \beta \\ \theta \\ \delta \\ \alpha \end{array}$ 
& $\begin{array}{c} 0.01 \\ 1.00 \\0.57 \\ 993.03 \end{array}$ $\begin{array}{c} \mathit{(0.21e^{-1})} \\ \mathit{(0.9e^{-2})} \\\mathit{(0.8e^{-1})} \\ \mathit{(121.08)} \end{array}$  
& $\begin{array}{c} 0.00 \\   1.00 \\ 0.94 \\ 0.55 \end{array}$ $\begin{array}{c} \mathit{(-)} \\ \mathit{(1.11)} \\ \mathit{(0.47)} \\ \mathit{(0.21)} \end{array}$ 
& $\begin{array}{c} 0.04 \\ 0.96\\  0.99 \\ 1.18 \end{array}$ $\begin{array}{c} \mathit{(1.1e^{-2})} \\\mathit{(0.7e^{-2})}\\ \mathit{(0.1e^{-2})} \\ \mathit{(2.8e^{-2})} \end{array}$ \\
\textbf{$\log\mathcal{L}$}  & \textbf{3874} & \textbf{4302} & \textbf{4360} \\
\textbf{BIC}  & \textbf{-7709} & \textbf{-8567} & \textbf{-8681} \\
\hline
\hline

\end{tabular}
\end{center}
\caption{\label{tab:par1} Estimation of the parameters of the copulas for $(H_s,W_s)$ couple. Standard deviations of the estimators are reported in italic.  }
\end{table}
\begin{table}[!ht]
\begin{center}
\begin{tabular}{cccc }

\hline
\hline
Model & Plackett & Gumbel & Clayton \\
\hline
One parameter  $\alpha$  & 9.27 $\mathit{ (3.8e^{-2})}$ &  0.51
$\mathit{(1.0e^{-5})}$ & 1.47 $\mathit{(3.5e^{-4})}$  \\
\textbf{$\log\mathcal{L}$}  & \textbf{5191} & \textbf{ 5528} & \textbf{4913} \\
\textbf{BIC} &  \textbf{-10373} &  \textbf{-11047} & \textbf{-9816} \\
\hline
Two parameters $\begin{array}{c} \delta \; (or \; \theta)  \\ \alpha \end{array}$ & 
 $\begin{array}{c} 0.78 \\ 15.17 \end{array}$  $\begin{array}{c}
\mathit{(1.0e^{-4})} \\ \mathit{(0.15)} \end{array}$ 
 & $\begin{array}{c} 0.85 \\ 0.46 \end{array}$ $\begin{array}{c}
\mathit{(1.5e^{-4})} \\ \mathit{(2.4e^{-5})} \end{array}$ 
 & $\begin{array}{c} 0.75 \\ 2.96 \end{array}$ $\begin{array}{c} \mathit{(7.0e{-3})}
\\ \mathit{(6.6e^{-2})} \end{array}$\\
\textbf{$\log \mathcal{L}$}  & \textbf{5305} & \textbf{ 5605} & \textbf{5453} \\
\textbf{BIC} & \textbf{-10591} & \textbf{-11192} & \textbf{-10886}\\
\hline
Three parameters $\begin{array}{c} \beta \\ \delta \; (or \; \theta)  \\ \alpha
\end{array}$ 
& $\begin{array}{c} 0.00 \\ 0.78 \\ 15.17 \end{array}$ $\begin{array}{c}
\mathit{(-)} \\ \mathit{(1.0e^{-4})} \\ \mathit{(0.15)}  \end{array}$  &
$\begin{array}{c} 0.19 \\   0.76 \\ 0.48 \end{array}$ $\begin{array}{c}
\mathit{(1.6e^{-2})} \\ \mathit{(1.5e^{-2})} \\ \mathit{(6.0e^{-3})} \end{array}$ &
$\begin{array}{c} 0.25 \\ 0.86 \\ 1.75 \end{array}$ $\begin{array}{c}
\mathit{(1.2e^{-2})} \\ \mathit{(6.0e^{-3})} \\ \mathit{(3.3e^{-2})}\end{array}$ \\
\textbf{$\log \mathcal{L}$}  & \textbf{5305} & \textbf{5682} & \textbf{5850} \\
\textbf{BIC}  & \textbf{-10581
} & \textbf{-11335} & \textbf{-11672} \\
\hline
\hline

\end{tabular} 
 
\end{center}
\caption{\label{tab:par2} Estimation of the parameters of the copulas for
$(H_s,T_p)$ couple. Standard deviations of the estimators are reported in italic.  }
\end{table}

~

\noindent
For the $(H_s,W_s)$ couples, the Clayton model with 4 parameters has the smallest  Bayesian Information Criterion (BIC). The estimated values of the parameters of this model give back the low asymmetry observed in the plotted copulas (Fig. \ref{fig:copulas}). Plackett's and Gumbel's models do not show the same behaviour. In these models only the global dependence parameter $\alpha$ is significant and equal to 6.76 for Plackett and 0.57 for Gumbel which is quite strong. Furthermore, the standard deviation of the estimators of the parameters are smaller for the Clayton model than for the other ones. Thus this model is shown here to be more flexible and lead to more robust estimators.

~

\noindent
The one-parameter Clayton model allows us to estimate the upper tail dependence at $0.57$ with a  standard deviation equal to $4.34e^{-3}$ and the Gumbel model at $0.52$ with a standard deviation equal to $7.33e^{-4}$. These values are sufficiently large to conclude that the  upper tail dependence is present.
The lower tail dependence is also present. Namely, we can test it with a likelihood ratio test, comparing  the four-parameter Clayton model to the three-parameter  Clayton model: 
here $-2(\log\mathcal{L}_{3}-\log\mathcal{L}_{4})=228$ and which is  significant for a $\chi^{2}$ statistic with one degree of freedom.

~

\noindent
For the $(H_s,T_p)$ couples where  the asymmetry is stronger, we choose the models defined by equation (\ref{eq:2par}) or equation(\ref{eq:clayta}) for reasons explained in section 1.  The introduction of the asymmetry parameter $\delta$ (or $\theta$ )  allows for better fitting  in all the  models.  Gumbel's and Clayton's models have a minimal dependence parameter greater than one which characterizes a low minimal dependence in the data.  It corresponds to sea states with low significant wave heights and short periods. As previously,  the Clayton model  with three parameters has the smaller BIC. It has been observed that the optimisation procedure used to fit the three-parameter Plackett model is very sensitive to the initialization. It is due to instability of the gradient close to the frontier of $\beta$ and also to the fact that the likelihood is flat for high levels of $\alpha$.
 The upper tail dependence is estimated at $.62$ ($\mathit{0.028}$) with the Clayton model and at $.56$  ($\mathit{6.73. e^{-3}}$) with Gumbel's model. Finally, the comparison of the two parameter-Gumbel model with the three-parameter Gumbel model gives a very significant likelihood ratio test  $-2(\log\mathcal{L}_{2}-\log\mathcal{L}_{3})=154$. Using the formula given in the 4th section of the appendix, we can estimate this lower tail dependence and it is equal to $\bar{\lambda}=0.63$.

\section{Conclusion}

In this paper, we have proposed  a procedure to construct a distribution model taking into account  a priori knowledge on the data. The main idea consists in transforming a basic copula such as the Plackett one  to better restore some features of the data distribution. Special attention is paid to asymmetry and extreme dependance. The transformed copulas have one to four parameters which are estimated by maximum likelihood. 

~

\noindent
The proposed procedure has been applied to sea state data. Two couples are considered. The first one $(H_s,W_s)$ presents maximal extreme dependence while the second one $(H_s,T_p)$ presents a clear asymmetry. Three basic copulas are studied: Plackett, Gumbel and Clayton. It is shown that the transformation of  basic copulas may improve the fitting of the distribution models, especially in the case of the Clayton model. However, we also observe that for the Plackett copula, which is more flexible than the Clayton one, the introduction of new parameters is not really useful.

~

\noindent
It is always difficult to have evidence of upper or lower dependence. Our method, by introducing a specific parameter devoted to the extreme dependence, allows us to test it and sometimes to estimate it from this parameter. The obtained estimator inherits the good properties of the ML-estimation (consistency and asymtotic normality).  Alternatively, it would be possible to use non parametric estimation of these indexes. But such estimation is often uncertain, linked to the visual appearance of the data, and  to the choice of a threshold (see Frahm et al. for more details \cite{Frahm}).

~

\noindent
The procedure described here could also be adapted to other assumptions on the copula,   for example   local dependence located outside the diagonal.

~

\noindent
Furthermore, we could have used a different  approach than Joe's generalization to 
obtain a model with minimal dependence.  Marsall and Olkin's procedure $
\hat{C}(u,v)=\int K(e^{-z\varphi(u)},e^{-z\varphi(v)} dG(z)
$ could also lead to such model.

~

\noindent
Finally, to model the assymetry, an alternative method could have consisted in defining an expression for the  boundary of the dataset and stipulating that the dependence is maximal in the vicinity of  this boundary using a procedure as explained by Rüschendorf \cite{Rusch}, starting from a function $\kappa(t)=t^{\beta}, \; 0<t<=1$ and modelling the boundary of the dataset.

\newpage 
\section*{\textbf{\large Appendix: Lower tail dependence}} 

\begin{enumerate}
\item The lower tail dependence of any  asymmetrized copula $\tilde{C}(u,v)=u^{1-\theta}C(u^{\theta},v)$ is zero.

Suppose that $C(u,v)$ is a symmetric copula with lower tail dependence such that $\lambda$ is greater than zero.
 Consider the asymmetrized copula $\tilde{C}(u,v)=u^{1-\theta}C(u^{\theta},v)$.
 $C(u,v)$  and $\tilde{C}(u,v)$ have positive dependence.

The positive dependence implies that
\[ u v  \le C(u,v) \le \min(u,v) ,\,\forall u,\,\forall v\]
where $uv$ corresponds to the independence copula and $\min(u,v)$ to the upper Fr\'echet bound.

In particular 
\[ u^{\theta+1} < C(u^{\theta},u) < u \]

Hence 
\[u < \frac{\tilde{C}(u,u)}{u} < \frac{u^{2-\theta}}{u}=u^{1-\theta}\]
When $u$ tends to $0$, the left and right hand terms of the inequality also tend to zero, as well as the middle term. And $\tilde{C}(u,v)$ has no lower tail dependence.

\item Lower tail dependence of Clayton's copula.

For Archimedean copulas, the lower tail dependence can be written 
\[ \lim_{u \to 0} \frac{C(u,u)}{u}=\lim_{u \to 0} \frac{\varphi^{-1}(2\varphi(u))}{u} \]
where $\varphi(.)$ is a decreasing function.
With $\varphi^{-1}(t)=(1+t)^{-\alpha}$, the $\lambda$ index of lower tail dependence of Clayton's copula is equal to $2^{-\alpha}$.

\item The lower tail dependence for  copula $\hat{C}(u,v)$ constructed with extreme dependence for the minimum is greater than Clayton's copula lower tail dependence.

We evaluate \[ \lim_{u \to 0} \frac{\hat{C}(u,u)}{u}=\\
\lim_{u \to 0} \frac{\varphi^{-1}(-\log C(e^{-\varphi(u)},e^{-\varphi(v)}))}{u} \]

where $C(u,v)$ is any copula with positive dependence.
The positive dependence implies that 
\[ uv  \le C(u,v) \le \min(u,v) \]

In particular, $ u^{2} < C(u,u) < u$. Since $\varphi(.)$ is decreasing, then $ e^{-\varphi(u)}$ is increasing according to $u$, so  this implies 
\[ e^{-2\varphi(u)} < C(e^{-\varphi(u)},e^{-\varphi(u)}) < e^{-\varphi(u)}\]
Taking minus the logarithm, we obtain 
\[2\varphi(u) > -log(C(e^{-\varphi(u)},e^{-\varphi(u)})) > \varphi(u)\]
Applying $\varphi^{-1}(.)$, which is also a decreasing function 
\[ \varphi^{-1}(2\varphi(u)) < \varphi^{-1}(-log(C(e^{-\varphi(u)},e^{-\varphi(u)}))) < \varphi^{-1}(\varphi(u)) \]

When $u$ tends to 0, $\frac{\varphi^{-1}(2\varphi(u))}{u}$ tends to $2^{-\alpha}$ and $\frac{\varphi^{-1}(\varphi(u))}{u}=1$, so that
\[ 2^{-\alpha}< \lim_{u \to 0} \frac{\varphi^{-1}(-\log C(e^{-\varphi(u)},e^{-\varphi(v)})))}{u} <1  .\]
This concludes the proof.

\item In some cases, we can evaluate the lower tail dependence $\bar{\lambda}_{\hat{C}}$ of $\hat{C}(u,v)$.

\begin{enumerate}

\item $C(u,v)$ has lower tail dependence $\bar{\lambda}_{C}$

In that case, $C(u,u)$ is equivalent to $\bar{\lambda}_{C}u$ when $u$ tends to 0.

\[\bar{\lambda}_{\hat{C}}=\lim_{u \to 0} \frac{\varphi^{-1}(-\log C(e^{-\varphi(u)},e^{-\varphi(u)}))} {u} \]
Let $t=e^{-\varphi(u)}$. When $u$ tends to $0$, $t$ also tends  to $0$. And
\[
\bar{\lambda}_{\hat{C}}=\lim_{t \to 0} \frac{\varphi^{-1}(-\log(C(t,t))}{\varphi^{-1}(-\log(t))} 
 =\lim_{t \to 0} \frac{\varphi^{-1}(-log(\bar{\lambda}_{C}t))}{\varphi^{-1}(-log(t))}=
\lim_{t \to 0} \frac{\varphi^{-1}(-log(t)-log(\bar{\lambda_{C}}))}{\varphi^{-1}(-log(t))}
 \]

That we can rewritten with $v=-log(t)$
\[ \bar{\lambda}_{\hat{C}}=\lim_{v \to \infty} \frac{\varphi^{-1}(v-\log(\bar{\lambda_{C}}))}{\varphi^{-1}(v)}
 \]
But since, $\varphi^{-}(t)=(1+t)^{-\beta}$ , we get 
 \[ \bar{\lambda}_{\hat{C}} =  \lim_{v \to \infty}(\frac{1+v-\log(\bar{\lambda_{C}}))}{1+v})^{-\beta}=\lim_{v \to \infty}(1-\frac{\log(\bar{\lambda_{C}}))}{1+v})^{-\beta}=1 \]

\item $C(u,v)$ has no lower tail dependence but $C(u,u)$ is equivalent to $\zeta u^{r}$, with $r>1$ when $u$ tends to 0.
With the same notation as in the preceding paragraph,
\[
\bar{\lambda}_{\hat{C}}=\frac{\varphi^{-1}(-\log(\zeta t^{r}))}{\varphi^{-1}(-\log(t))}=
\lim_{t \to 0} \frac{\varphi^{-1}(-r\log(t)-\log(\zeta))}{\varphi^{-1}(-\log(t))}=\lim_{v \to \infty} \frac{\varphi^{-1}(rv-\log(\zeta))}{\varphi^{-1}(v)}
 \]

This is the case for Gumbel's copula. Indeed,

\[C_{G}(u,u)=u^{2^{\alpha}}, \;  0<\alpha<1 .\]
Hence $C_{G}(u^{\theta},u)=u^{(\theta^{\frac{1}{\alpha}}+1)^{\alpha}}$ and 
$\tilde{C}_{G}(u,u)= u^{1-\theta}u^{(\theta^{\frac{1}{\alpha}}+1)^{\alpha}}$

Therefore, choosing $r=1-\theta+(\theta^{\frac{1}{\alpha}}+1)^{\alpha}$
\[ \bar{\lambda}_{\hat{C}_{G}}= \lim_{v \to \infty} \frac{\varphi^{-1}(rv)}{\varphi^{-1}(v)}=\lim_{v \to \infty} (\frac{1+rv}{1+v})^{-\beta}=r^{-\beta}. \]

\end{enumerate}
\end{enumerate}

\bibliographystyle{elsarticle-num}

\end{document}